# On-Chip Communication Network for Efficient Training of Deep Convolutional Networks on Heterogeneous Manycore Systems

Wonje Choi, *Student Member, IEEE, Karthi Duraisamy, Student Member, IEEE,* Ryan Gary Kim, *Member, IEEE,* Janardhan Rao Doppa, *Member, IEEE*, Partha Pratim Pande, *Senior Member, IEEE,* Diana Marculescu, *Fellow, IEEE*, and Radu Marculescu, *Fellow, IEEE*

Abstract— Convolutional Neural Networks (CNNs) have shown a great deal of success in diverse application domains including computer vision, speech recognition, and natural language processing. However, as the size of datasets and the depth of neural network architectures continue to grow, it is imperative to design high-performance and energy-efficient computing hardware for training CNNs. In this paper, we consider the problem of designing specialized CPU-GPU based heterogeneous manycore systems for energy-efficient training of CNNs. It has already been shown that the typical on-chip communication infrastructures employed in conventional CPU-GPU based heterogeneous manycore platforms are unable to handle *both* CPU and GPU communication requirements efficiently. To address this issue, we first analyze the on-chip traffic patterns that arise from the computational processes associated with training two deep CNN architectures, namely, LeNet and CDBNet, to perform image classification. By leveraging this knowledge, we design a hybrid Network-on-Chip (NoC) architecture, which consists of both wireline and wireless links, to improve the performance of CPU-GPU based heterogeneous manycore platforms running the above-mentioned CNN training workloads. The proposed NoC achieves **1.8**× reduction in network latency and improves the network throughput by a factor of **2.2** for training CNNs, when compared to a highly-optimized wireline mesh NoC. For the considered CNN workloads, these network-level improvements translate into **25**% savings in full-system energy-delay-product (EDP). This demonstrates that the proposed hybrid NoC for heterogeneous manycore architectures is capable of significantly accelerating training of CNNs while remaining energy-efficient.

## 1 Introduction

Deep learning techniques have seen great success in diverse application domains including speech processing, computer vision, and natural language processing [1]. While the fundamental ideas of deep learning have been around since the mid-1980s [2], the two main reasons for their recent success are: 1) the availability of large-scale training data; and 2) advances in computing hardware to efficiently train large-scale neural networks using this data.

*Deep learning* refers to a class of machine learning algorithms, where the goal is to train a non-linear function approximator represented as a neural network architecture with multiple layers of neurons by using input-output pairs of training data. In this work, we consider a special class of deep learning architectures, namely *Convolutional Neural Networks (CNNs)*. CNNs are employed to learn hierarchical representations of different modalities of raw data such as text, images, audio, and video.

CNNs are composed of several layers. Essentially, each CNN layer implements an operator which takes a three-dimensional data tensor as input and transforms it into a resultant three-dimensional data tensor. These CNN layers are then cascaded to implement a composite operation, where the first layer corresponds to raw input data and the result of the last layer corresponds to the predicted output. For example, in the digit classification task, the result of the last layer is a vector that contains output probability scores of different digit labels for a given input image.

Mathematically, a CNN is a non-linear function whose parameters are represented by the *weights* or *filters* present in the intermediate CNN layers. The values of these weights are commonly learnt using an iterative training algorithm called *backpropagation* (short for "backward propagation of errors") [2]. At each iteration of the backpropagation algorithm, the CNN first computes the predicted output by forwarding a sample input through the network, employing the current set of weights (forward pass). Then, the gradient of the prediction error is computed and passed backwards through the network (backward pass). Finally, the weights of the network are updated using the gradient(s) from the prediction error through a variant of stochastic gradient descent (SGD) optimization [2].

Training CNNs is a highly data- and compute-intensive task. It involves complex vector and matrix computations at each layer. Simultaneously, CNN training also exhibits a high degree of data parallelism. By exploiting this parallelism, the above-mentioned operations associated with training CNNs can be significantly accelerated using GPU cores [3]. Indeed, GPU-based manycore platforms are the most preferred design choice to implement various neural network applications across both industry and academia. Caffe and TensorFlow are two key examples of frameworks for deep learning targeted at GPU-based systems [4][5]. Moreover, NVIDIA has recently introduced the DGX-1, a GPU-based supercomputer targeting Artificial Intelligence and Deep Learning applications and the Jetson TX2, a heterogeneous embedded system architecture, to efficiently handle deep learning applications.

Aside from the high data parallelism, CNN training also involves high volumes of message transfers between the CPUs and GPU-accelerators, mainly due to the forwarding and storing of data between adjacent CNN layers [6][7]. In a discrete GPU system, the communication between the CPUs and GPUs is carried out via off-chip interconnects (*e.g.*, PCIe) that exhibit high data-transfer latency and power consumption [6][8][9]. A heterogeneous single chip multiprocessor (CMP) where the CPUs and GPUs are interconnected through the on-chip network will avoid such expensive off-chip data transfers. Hence, heterogeneous CMPs promise to be the most suitable option for efficient execution of CNN applications.

We note that, conventional data centers and high-performance computing (HPC) clusters are employed to solve deep learning applications. However, the design of data centers and HPC clusters is dominated by power, thermal, and area constraints. Hence, we envision a Datacenter-on-Chip (DoC) architecture specifically targeting deep learning applications where the entire system, or a large part thereof, can be designed using heterogeneous manycore-based single-chip architectures. It is well understood that with this massive level of integration, traditional Network-on-Chip (NoC) architectures, *e.g.*, mesh, tree, ring, cannot provide a scalable, low-latency, and energy-efficient communication backbone, which is essential for solving the deep learning problems targeted in this work [10]. On the other hand, wireless NoC (WiNoC) can achieve energy–efficient and low-latency communication infrastructures for massive manycore chips [11][12]. It has been shown that the mm-wave based WiNoC outperforms conventional wireline NoC architectures in terms of achievable bandwidth and energy dissipation [13]. Consequently, inspired by the successes of this WiNoC, in this work:

- We explore heterogeneous (*i.e.*, combination of CPUs and GPUs) systems combined with hybrid (*i.e.*, combination of wired and wireless links) NoC architectures for deep learning.
- We present a generic design methodology that can be instantiated for any combination of neural networks and application domains.
- We evaluate the efficacy of our design methodology using two real-world CNN architectures, namely LeNet and CDBNet, performing image classification [14][15].
- We demonstrate through rigorous experiments that the proposed NoC achieves 25% full-system Energy Delay Product (EDP) reduction while accelerating the training of CNNs by 15%.

The remainder of the paper is organized as follows. In Section 2, we present some of the related works and highlight our novel contributions. In Section 3, we provide an overview of the CNNs along with the computational perspective of CNN training process. In Section 4, we present our generic heterogeneous NoC design methodology. In Section 5, we discuss the experimental results to demonstrate the efficiency of the proposed NoC over the traditional wireline counterparts. Finally, Section 6 concludes the paper by summarizing the findings and outlining future directions of this work.

## 2 Related Work

CNN technology has seen success in a wide-range of applications including computer vision (*e.g.*, image classification [16] and video classification [17]); speech processing [18]; natural language processing (*e.g.*, text classification [19] and learning semantic representations for web search [20]); drug design [21]; and to represent policies and value functions for driving the Monte Carlo tree search approach employed in AlphaGO [22].

NoC-enabled homogeneous CMP architectures targeting neuroscience applications have already been explored. For instance, a massively parallelized CMP platform incorporating a customized NoC architecture was used to implement spiking neural networks [23]. Multicast-aware mesh NoC architectures have been proposed for reconfigurable neural networks [24][25]. However, due to the highly data-parallelizable nature of the neural networks, these applications have already been demonstrated to be more efficient on discrete GPU systems rather than traditional multi-CPU CMPs [3][26]. The design of a commodity off-the-shelf system for HPC targeting deep learning was proposed recently [27]: a cluster of GPU servers with Infiniband interconnects and a message-passing interface show promise over large CPU-only based systems. Recently, the architecture of a machine learning "supercomputer" [26] that achieves higher performance and lower energy dissipation than a modern GPU-based system was also proposed. The system relies on a multichip design, where each node is significantly cheaper than a typical GPU while achieving comparable or higher number of operations per unit time in a smaller package. A recent study characterized the performance of two popular CNN architectures (*i.e.,* LeNet, CDBNet) on various embedded platforms including FPGAs, DSPs, and GPUs [7]. The main limitation of [7] is that it mainly focuses on optimizing the software parallelization and mapping, without a detailed on-chip network analysis required for an efficient many-core-based implementation.

Prior works on discrete GPU platforms have focused on improving the system performance by enhancing their NoC architectures [28][29][30]. In a GPU system, processes executed in each GPU core are usually independent of the other GPUs' processes, resulting in low inter-GPU communication [28]. Typically, GPUs only communicate with a few shared memory controllers (MC), causing a many-to-few traffic pattern (*i.e.,* many GPU cores communicating with a few MCs) [28][29][30]. In this case, MCs can potentially become traffic hotspots and lead to performance bottlenecks. Prior research demonstrated that suitable placement of the MCs can help alleviate the associated traffic congestion [28][29]. To prevent traffic imbalance among the links, a checkerboard mesh NoC with a suitable routing strategy was recently proposed [28]. Applications running on GPUs tend to exhibit an asymmetric traffic pattern, where the amount of communication from L1 to MC is usually lower than that from MC to L1 due to memory coalescing in GPU architectures [30]. This asymmetric nature further worsens the performance bottlenecks that are exhibited under the many-to-few traffic pattern. To overcome this issue, an asymmetric virtual channel partitioning and monopolization technique for discrete GPU NoCs was proposed [29]. The advantage of using a clustered mesh NoC (with four L1s per cluster) over non-clustered mesh and crossbar architectures (*i.e.*, all L1s in a single cluster) for discrete GPU systems was demonstrated [30].

As explained above, training CNNs involves heavy CPU-GPU communication that is best suited for a NoC-enabled heterogeneous CPU-GPU CMP platform rather than a traditional discrete-GPU system with expensive off-chip CPU-GPU data transfers [9]. Due to the differences in the thread-level parallelism of CPUs and GPUs, the NoC employed for heterogeneous systems is expected to handle traffic patterns with varying Quality of Service (QoS) constraints [31]. CPUs are highly sensitive to the memory access times and hence, communications involving CPUs require low-latency data exchanges. On the other hand, GPU communication demands high bandwidth [31]. Modern NoC designs for discrete GPU systems typically attempt to only maximize the overall bandwidth of the system. Consequently, these NoC designs are unsuitable for heterogeneous CMP architectures incorporating multiple CPUs and GPUs on the same die. It has been shown that the shared memory resources in a heterogeneous system are often monopolized by the GPUs, leading to significant degradation in CPU memory access latency and high execution time penalties [32]. Hence, an efficient on-chip network designed for heterogeneous CMPs should balance the inter-core traffic and comply with different QoS standards required for both CPU and GPU communications.

The NoC design for CPU-GPU heterogeneous systems have yet to be studied thoroughly. A system-level discussion regarding the NoC design for CPU-GPU heterogeneous architectures was presented recently [31]. However, this work only considered a ring interconnect that is known to be inefficient for large-scale systems. A wireless NoC was proposed for heterogeneous systems [33]. However, this work does not provide any detailed analysis of the on-chip traffic patterns and has not considered the nature of traffic patterns that is inherent in CPU-GPU heterogeneous platforms (*e.g.*, many-to-few data exchanges) while designing the wireless NoC. A virtual channel partitioning scheme was proposed to achieve low-latency CPU-related memory accesses in the presence of largely latency-insensitive GPU communication [34]. However, this strategy can affect the GPU throughput due to the partitioning of physical resources. As we will explain later, our work employs dedicated wireless links for CPU-MC communication to avoid this network contention and hence, ensures a high GPU-MC throughput.

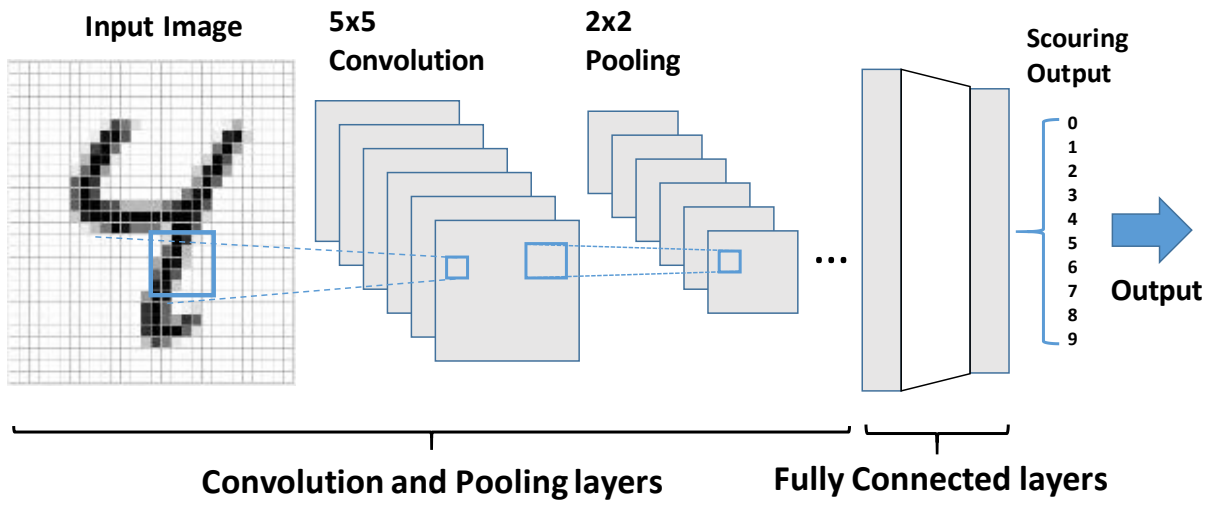

**Fig.1. High-level overview of the CNN architecture for digit classification task.**

Thus far, most of the NoCs targeting discrete GPU systems are based on conventional wired NoC architectures. The achievable performance benefit from the proposed strategies is restricted due to the inherent limitations associated with these NoCs. Traditional wireline NoCs (such as mesh) use multi-hop, packet-switched communication that lead to high network latencies [10]. To overcome these limitations, small-world network-inspired wireless NoC architectures have been proposed [11][12]. Indeed, by employing a few long-range wireless shortcuts, these architectures enable low latency communication even among the computing cores that are physically far apart. Previous works [35][36][37] have already investigated the feasibility of the on-chip wireless communication. The viability of on-chip wireless communication has been demonstrated through prototypes [13]. In addition, a recent study on emerging on-chip interconnects concluded that the Radio-Frequency (RF) links, e.g., mm-wave wireless and surface-wave interconnects, are more power and cost efficient than on-chip optical links [38]. Between the two RF interconnects, on-chip wireless technology is more mature and fully CMOS compatible [12][38].

While designing NoCs enabled by wireless links, it should be noted that the number of wireless interfaces (WIs) that can be added in a NoC is usually limited by number of factors. A couple of such factors are: (1) the nature of the medium access control (MAC) protocol employed, and (2) the area and power overhead allowed for the addition of WIs. Hence, it is necessary to use innovative system architecture designs to harvest the full potential of wireless-enabled NoCs [39]. Towards this goal, in this work, we improve the state-of-the-art by presenting a hybrid (wireline + wireless) on-chip interconnection architecture that can meet the communication demands uniquely present in a heterogeneous CMP platform consisting of both CPU and GPU cores.

# 3 Deep Convolutional Neural Networks

## 3.1 Overview

CNNs usually perform prediction/classification tasks (*e.g.*, finding the output probability scores of different class labels for a given input image) where the inputs and the outputs are fixed size vectors. A typical CNN architecture consists of a stack of alternating convolutional and pooling layers, followed by a fully-connected layer at the end. Fig. 1 is a high-level conceptual diagram illustrating a CNN architecture performing *hand-written digit classification*.

**Convolutional layer:** Convolutional layers are the key building blocks of a CNN architecture and are comprised of sets of learnable filters (or kernels) that aim to transform a low-level representation (*e.g.*, raw pixels of an image) into a high-level representation (*e.g.*, edges/sections in the image). During the forward pass, each filter is convolved by sliding over the input to generate a feature map. Convolutional layer computations are commonly followed by an element-wise non-linear activation function (*e.g.*, Rectified Linear Unit (ReLU) or sigmoid).

**Pooling layer:** Typically, a pooling layer is employed between two convolutional layers. It aggregates the input representation succinctly using a *max* or *average* operator. The purpose of this aggregation is to reduce the volume of computations and parameters associated with the network and to avoid over-fitting. Pooling operations can also be seen as a spatial down-sampling of the input 3D tensor.

**Fully connected layer**: At the end of the convolutional neural networks, all the outputs of the second to last layer (usually a convolutional or a pooling layer) are connected to the entire output vector, constituting a fully connected layer. Thus, the size of this layer, *i.e.*, the number of neurons, is equal to the number of output classes, *e.g.*, ten in digit classification.

**Network weights:** The weights or parameters of a given CNN architecture, *W*, correspond to the different filters at the convolutional layers and the weights of the fully connected layer. It should be noted that the pooling layer does not have any parameters.

**Prediction with given weights**: Given an input image $x = [x_{11}, x_{12}, \dots]$ and network weights $W = [w_{11}, w_{12}, \dots]$, we propagate the input forward through the network sequentially through all the layers to compute the output of the network: scores for each class; the highest scoring class is selected as the prediction.

## 3.2 CNN Training Datasets

In this work, we consider two of the most widely used image recognition datasets, namely MNIST and CIFAR-10 [40][41], and train suitable CNN architectures to perform the image classification tasks associated with these two datasets. The MNIST dataset consists of handwritten digits (in greyscale) with 60,000 training and 10,000 testing samples. CIFAR-10 is an object recognition dataset consisting of color images containing one of 10 different object classes with 50,000 training and 10,000 testing examples.

We have adopted the LeNet architecture [14] to solve the handwritten digit classification task using the MNIST database. This LeNet architecture consists of an input layer followed by two sets of alternating convolutional and max-pooling layers. These layers are then followed by a convolutional layer, a fully connected layer, and finally, the output layer. We have employed the Convolutional Deep Belief Neural Network (CDBNet) architecture from ConvNet [15] to perform an image classification task using the CIFAR-10 dataset. CDBNet consists of three convolution layers followed by a local normalization layer, a max-pooling layer, two average-pooling layers, and one fully connected layer. We have shown the detailed layer configuration of both LeNet and CDBNet in Table 1.

The object classification task of CIFAR-10 is more complex than the MNIST image recognition task as it involves colored images and ten different object classes. Hence, compared to LeNet that uses only 16 filters, CDBNet is more complex and employs 64 filters. Moreover, unlike LeNet that uses max-pooling method for all pooling layers, CDBNet employs the average-pooling method for its second and third pooling layers.

**Table 1. Layer configurations for LeNet and CDBNet. Each entry includes the dimensions of the layer's input and the dimensions of the kernel (excluded for the input column).**

| Dataset (CNN) | Input | 1st conv | 2nd conv | Last conv |
|---|---|---|---|---|
| MNIST (LeNet) | 33x33x1 | 29x29x16<br>5x5 kernel | 11x11x16<br>5x5 kernel | 1x1x128<br>5x5 kernel |
| CIFAR-10 (CDBNet) | 31x31x3 | 31x31x32<br>5x5 kernel | 15x15x32<br>5x5 kernel | 7x7x64<br>5x5 kernel |

## 3.3 Training CNNs

Conventionally, the weights of a CNN are learned from a set of training data using the backpropagation algorithm. Given an input image $x$ and weights $W$, we propagate the input forward through the network to compute the predicted output $\hat{y}$. Given a set of $T$ training input-output pairs $(x_k, y_k)$, the weights of the network are learnt such that the error of the predicted outputs of the neural network (as described above) is minimized.

Typically, an iterative optimization training algorithm (backpropagation) is used to learn the network weights. At each iteration, for each input example $x_k$ from the training data, we make a prediction $\hat{y}_k$ by making a forward pass through the network using the current network weights (as described above); we then compare the predicted output $\hat{y}_k$ with the correct output $y_k$. If there is an error, we attempt to correct it by making a backward pass through the network, compute the gradient vector $\delta_k$, and perform a weight update using the cumulative gradient as follows:

$$\delta = \sum_{k=1}^{T} \delta_k \quad (1)$$

$$W = W + \eta\delta \quad (2)$$

where $\eta$ is the learning rate and $W$ is the weight vector corresponding to all network weights.

It should be noted that during both the forward and backward passes, a lot of data parallelism exist in the computations. In fact, all the neurons in a CNN layer perform the same computation with different input and filter parameters. Hence, this data parallelism can be easily exploited using hardware accelerators (such as GPUs) to greatly accelerate CNN training.

For both forward and backward passes, the convolutional layers contain the most computationally expensive operations. Each convolution operation requires multiple dot products of the filter values and the sectional input vectors. Standard implementations of the convolutional layer mainly involve matrix operations and therefore, leverage efficient libraries such as Basic Linear Algebra Subprograms [42] for GPGPU-based systems. The activations at fully connected layer can also be computed using matrix multiplication.

# 4 NoC For Heterogeneous Platform

In CPU-GPU heterogeneous manycore architectures, the NoC primarily handles *many-to-few* communication patterns. Each processing core (either a GPU or a CPU) in the system maintains its own L1 cache. The L1 caches mainly exchange data with a limited number of MC blocks and the volume of inter-GPU data-exchanges is negligible [28][30]. Each MC incorporates a Last Level Cache (LLC) and a mechanism to access the main memory. Fig. 2 illustrates a CPU-GPU heterogeneous architecture[1] with four CPUs, 28 GPUs, and four MCs. As shown in Fig. 2, in order to perform efficient many-to-few communication, several GPUs can be attached to a single router (concentration) [30].

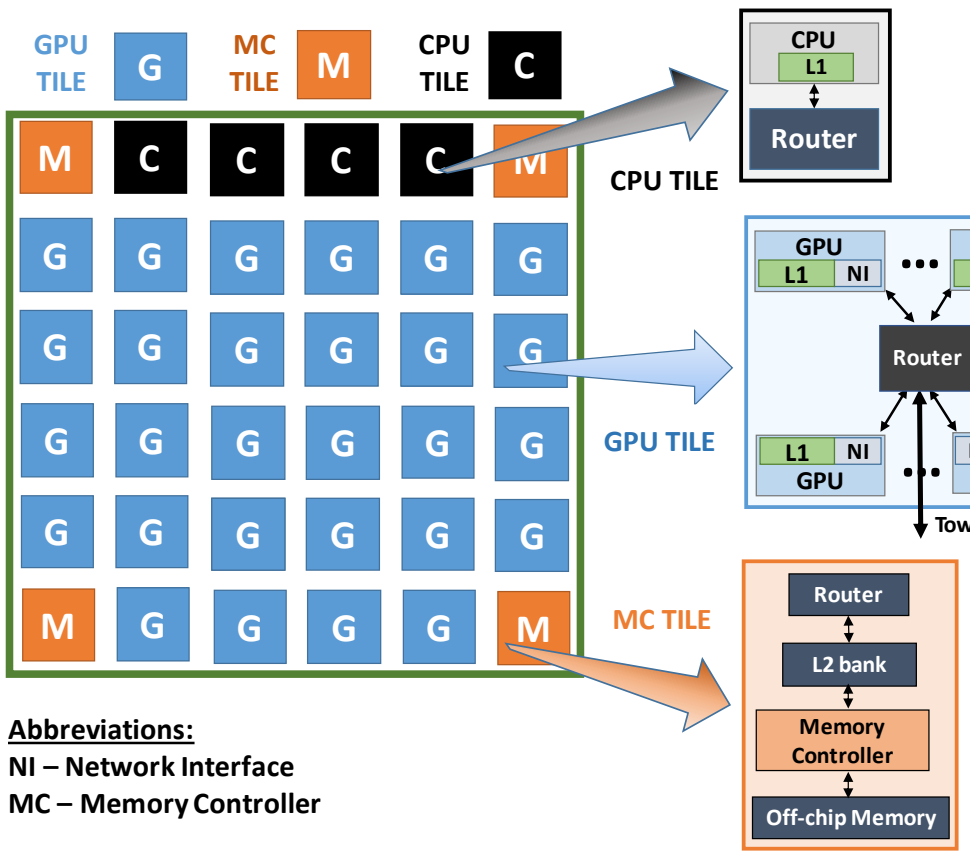

**Fig. 2. Illustration of a heterogeneous architecture incorporating CPUs, GPUs, and MCs. Each tile contains a network router enabling NoC interconnections. Because of L1 concentration, GPU tiles incorporate 4 GPU cores and 4 L1s in each tile.**

In a heterogeneous manycore platform, the QoS requirements for the GPU-MC and CPU-MC communications vary from each other. The CPU-MC communications are primarily latency-sensitive, while the GPU-MC communications are more throughput-sensitive [31]. In CPUs, waiting for memory accesses lead to stalled processor cycles resulting in execution time penalties. On the other hand, GPU's low-cost context switching makes them less susceptible to GPU-MC communication latency. However, each GPU core consists of multiple thread execution units, requiring large streams of data exchanges between the GPU and MC, leading to high throughput requirements [31]. Considering the above-mentioned facts, a NoC designed for heterogeneous CPU-GPU systems must be optimized to ensure that the CPU-MC communication latency is minimized while the overall NoC throughput is maximized. Hence, in this work, we consider these two objectives and jointly optimize them while designing the overall interconnection architecture.

### 4.1 Mesh NoC

It was shown that on a mesh NoC for handling *many-to-few* communication patterns, placing the MCs closer to the middle rather than along the chip edges reduce traffic congestion in the links and ensures better overall NoC throughput [29]. However, this mesh NoC was designed for a homogeneous GPU-based system. In the case of heterogeneous architectures, which is the focus of this work, we need to consider the placement of the CPU cores in addition to the placement of the MCs to achieve both low latency CPU-MC communication and high NoC throughput. However, as we will show in Fig. 8, even in a mesh optimized for both CPU and MC placements, there exist a few links that are heavily utilized when compared to the rest of the links present in the NoC. During high traffic, such links will become bandwidth bottlenecks, negatively affecting the overall system performance. The presence of these bandwidth bottlenecks can be attributed to the many-to-few communication pattern and the multi-hop nature of the mesh NoC architecture, which leads to high traffic aggregation in the intermediate routers and links.

To address the inherent multi-hop nature of mesh NoCs, the design of wireless NoCs (WiNoCs) have been proposed [11][12]. The salient feature of the WiNoC is that the wireless links establish single-hop shortcuts between physically distant cores, thereby improving the hop-count and subsequently, the latency, throughput, and energy dissipation of the whole system. Hence, we aim to design a customized low hop-count WiNoC targeted for the heterogeneous architecture under consideration.

### 4.2 Proposed Hybrid NoC

In this work, we propose a NoC architecture comprising both wireline and wireless links customized for the CPU-GPU heterogeneous computing platform. We call this proposed NoC a *Wireless-enabled Heterogeneous NoC (WiHetNoC)*. We intend to use dedicated single-hop wireless links between the CPUs and MCs. It has been already demonstrated in [12][35][39] that for multi-hop on-chip communication, mm-wave wireless links can achieve a lower EDP compared to a wireline link. Hence, we intend to use dedicated single-hop wireless links between CPUs and MCs. The GPU-MC communication is handled through a combination of wireline and wireless links that are tailored to the many-to-few traffic pattern. Aside from enabling low latency CPU-MC data exchanges, the use of dedicated wireless links for CPU-MC communication makes the WiHetNoC design agnostic of the CPU and MC placements, instantly fulfilling the CPU QoS requirements. This is because the wireless links guarantee direct single-hop communication regardless of the physical distance between the transceivers as long as they are within the communication range. We can then focus on fulfilling the GPU QoS requirements by maximizing the throughput of the WiHetNoC. We begin by creating the underlying wireline connectivity to optimize the GPU QoS through multi-objective optimization (MOO). Then, through careful placement of wireless interfaces we create the final WiHetNoC architecture. Fig. 3 shows the overall design flow to create the WiHetNoC.

#### *4.2.1 Problem Formulation.*

In this section, the main problem we examine is the optimization of the NoC link placement in CPU-GPU heterogeneous platforms that run deep learning applications. To identify potential bandwidth bottlenecks, we compute the expected utilization of each link in the NoC.

For a NoC with $R$ routers and $L$ links, we can find $U_k$,

[1]It should be noted that the tile locations in this heterogeneous architecture figure are just for illustration purposes. They are not optimized for any specific performance metric.

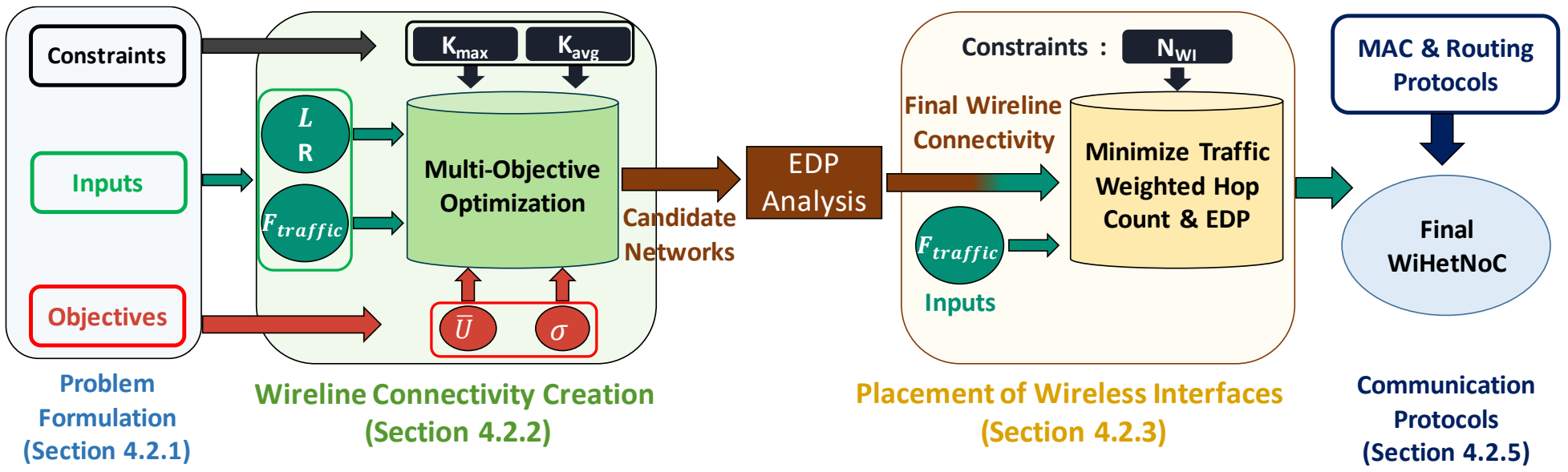

**Fig. 3. WiHetNoC design flow. Required input parameters and the objectives optimized in the creation of WiHetNoC are also shown.**

the expected utilization of any link $k$ by using the following equation:

$$U_k = \sum_{i=1}^{R}\sum_{j=1}^{R} f_{ij} p_{ijk}\,,\; p_{ijk} = \begin{cases} 1, & if\ i,j\ communicate\ along\ link\ k \\ 0, & otherwise \end{cases} \quad (3)$$

where $p_{ijk}$ indicates which links are used for the communication from router $i$ to router $j$. The $p_{ijk}$ value can be found by using the network connectivity and NoC routing protocol. The value $f_{ij}$ denotes the frequency of interaction between routers $i$ and $j$. It should be noted that these $f_{ij}$ values are not specific to any neural network (NN) layer of the NN architecture. Rather, the $f_{ij}$ values are a representation of the many-to-few traffic pattern that is associated with the considered heterogeneous system.

In our WiHetNoC, we establish the network connectivity such that both the mean link utilization ($\bar{U}$) and the standard deviation among the link utilizations ($\sigma$) are minimized. $\bar{U}$ and $\sigma$ can be determined as below,

$$\bar{U} = \frac{1}{L}\sum_{k=1}^{L} U_k = \frac{1}{L}\sum_{k=1}^{L}\left(\sum_{i=1}^{R}\sum_{j=1}^{R} f_{ij} p_{ijk}\right) = \frac{1}{L}\sum_{i=1}^{R}\sum_{j=1}^{R}\left(f_{ij}\sum_{k=1}^{L} p_{ijk}\right) = \frac{1}{L}\sum_{i=1}^{R}\sum_{j=1}^{R} f_{ij} h_{ij} \quad (4)$$

$$\sigma = \sqrt{\frac{1}{L}\sum_{k=1}^{L}(U_k - \bar{U})^2} \quad (5)$$

Here, $h_{ij}$ denotes the minimum number of inter-router hops between routers $i$ and $j$ with the given network configuration. Intuitively, $h_{ij}$ is equal to the total number of links used in the communication between router $i$ and router $j$.

As it can be observed from the above equations, $\bar{U}$ is directly proportional to the traffic weighted hop count (given by $\sum_i^R \sum_j^R f_{ij} h_{ij}$). Thus, minimizing $\bar{U}$ also minimizes the inter-router hop count, leading to high network throughput. On the other hand, minimizing $\sigma$ ensures that the link utilizations in the WiHetNoC are well-balanced and free of bandwidth-bottlenecks.

There are several constraints associated with establishing the network connectivity, and hence, the overall problem can be formulated as a constrained optimization problem with the following parameters:

| | | |
|---|---|---|
| Inputs | : $f_{ij}$, R, L | |
| Outputs | : Network connectivity, $d^*$ | |
| Objectives | : $d^* = \arg min_{d \in D}\, f(\bar{U}(d), \sigma(d))$ | (6) |
| Subject to | : $\frac{1}{R}\sum_{r=1}^{R} k_r \leq k_{avg}$ | (7) |
| | : $k_r \leq k_{max}, \forall r$ | (8) |
| | : $fullyConnected(d^*) == true$ | (9) |

where $D$ is the set of all possible network configurations, $k_{avg}$ and $k_{max}$ are average and maximum router port count, $k_r$ is the router port count for router $r$, and $d^*$ is the connectivity obtained from optimizing $\bar{U}$ and $\sigma$. $fullyConnected(d^*)$ returns true if there exists a path between all pairs of nodes in the given network $d^*$. This ensures each node can communicate with any other node.

### *4.2.2 Optimizing Wireline Network Connectivity*

Since minimizing both $\bar{U}$ and $\sigma$ together is a MOO problem, and the full search space of (6), before constraints, is $\binom{R*(R-1)}{L}$, where $R$ is the number of routers (one per tile) and $L$ is the number of links. Searching this entire space for a realistic NoC involving many cores (64 cores for this work) to determine the global optima is clearly intractable. Additionally, ILP, MILP or any other LP solvers are not suitable for this problem since nether optimization objective is a linear function (average link utilization and variance, Eqns. 4 and 5). In this work, we employ Archived Multi-Objective Simulated Annealing (AMOSA) [43] to solve (6). AMOSA is a simulated annealing (SA) based algorithm in which the optimization process is guided with the help of an archive of solutions. AMOSA is a widely-used MOO algorithm that can find a near-optimal Pareto front for many applications [42]. Hence, we considered it as a suitable candidate MOO algorithm to solve the problem undertaken here. However, any other MOO methodology (*i.e.*, genetic and evolutionary) can be employed to perform this optimization. Exploring such different MOO optimizations may help us in identifying the technique that has the fastest convergence time with comparable quality to AMOSA (identifying the fastest MOO algorithm may be essential to design large-scale heterogeneous systems). However, in this work, we mainly focus on creating a framework to design the wireless-enabled NoC for heterogeneous systems (using CNN training as a case study) by formulating and analyzing the necessary objectives and constraints. Hence, the exploration of the fastest MOO for heterogeneous architecture design is left for future work.

In AMOSA, during each optimization step, a perturbation is created in one of the archived solutions to generate a new configuration. Depending on the comparative quality of this new configuration over all solutions in the archive, AMOSA then updates the archive. Thus, on completion of AMOSA, we obtain a set of archived candidate configurations.

Since there is no structure enforced in the NoC connec-tivity while using AMOSA (*e.g.*, forced neighborhood connectivity) and the traffic characteristics show highly imbalanced communication, the optimized NoC would generally follow an irregular connectivity. Therefore, we need to follow certain physical restrictions while establishing the link connectivity in this irregular WiHetNoC. First, we limit the average number of inter-tile communication ports per router ($k_{avg}$) in (7) to four so that the WiHetNoC does not introduce any additional router port overhead when compared to a conventional mesh. Next, we need to restrict the maximum number of ports in a router ($k_{max}$) in (8) so that no router becomes unrealistically large. The MCs in a heterogeneous NoC are traffic hotspots with heavy volumes of incoming and outgoing messages. Increasing $k_{max}$ allows the number of router ports attached to an MC to increase, and hence, improves the MC router bandwidth. However, high $k_{max}$ values can lead to large routers, which result in high network energy consumption. Moreover, large routers make the whole system highly vulnerable to failures. Consequently, we consider a $K_{max}$ range of four to seven. For each $K_{max}$ value, we create a candidate network set through AMOSA that minimizes the mean link utilization and the standard deviation among the link utilizations. Then, among these candidate networks, we choose the network with the lowest EDP as the optimum wireline connectivity for the WiHetNoC (experimental results are shown later in Section 5).

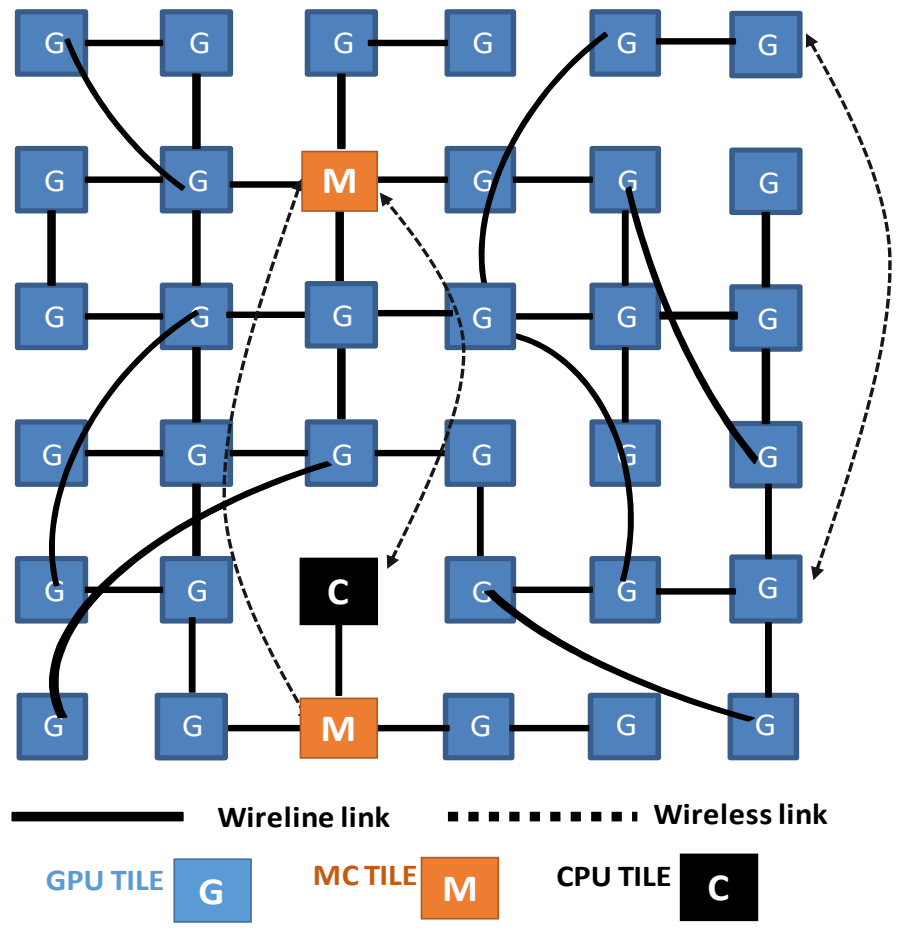

**Fig. 4. Illustration of WiHetNoC Connectivity.**

### 4.2.3 Wireless Link Placement

As we illustrate in Fig. 4, the WiHetNoC designed following the methodology described in Section 4.2.2 generally contains several long wireline interconnections. As these links are extremely costly in terms of power and delay, we employ wireless links operating in millimeter (mm)-wave and sub-THz range of 10-220 GHz to connect the routers that are separated by long distances. Since the CPUs are latency sensitive, CPU-MC communication is handled using dedicated wireless links.

In practice, depending upon the available wireless resources, we can only make a limited number of the longest links wireless, while the other links need to remain wireline. We use five non-overlapping channels centered around 30, 60, 90, 140 and 200 GHz. Each channel sustains a data rate of 16 Gbps for a communication range of at least 20 mm. It should be noted that by using the current CMOS technology both the frequency range and number of channels could be increased further. However, the proposed design methodology presented in this work is oblivious to these physical design parameters, *i.e.*, the number of wireless channels can be increased or decreased without modifying the proposed algorithm. Using these five channels we overlay the wireline connectivity with the wireless links such that a few routers get an additional wireless port. The wireless ports have a wireless interface (WI) tuned to one of the five different frequency channels.

Given the total number of WIs allowed ($N_{WI}$), we use a WI placement strategy that focuses on minimizing traffic-weighted hop-count [44]. Following this methodology and by varying $N_{WI}$, we find both the optimum number of wireless interfaces and the best locations in the WiHetNoC. The optimum value of $N_{WI}$ is discussed later in experimental results and analysis section (Section 5).

### 4.2.4 Components of the Wireless Interface

The two principal components of a wireless interface are the antenna and the transceiver. WiHetNoC uses a metal zigzag antenna that has been demonstrated to provide the best power gain with the smallest area overhead [12]. A detailed description of the transceiver circuit is out of the scope of this paper. However, the transceivers were designed following the principles described in [11]. The wireless interface is completely CMOS compatible and no new technology is needed for its implementation. In the 28nm technology node, for data rates of 16 Gbps, wireless links dissipate 1.3 pJ/bit (18 mW power) over a 20mm communication range and each WI occupies 0.25$mm^2$ area.

### 4.2.5 Communication Protocols

The proposed NoC principally has an irregular application-specific topology and requires a topology-agnostic routing method. We follow the Adaptive Layered Shortest Path (ALASH) routing methodology [11]. ALASH is built upon the layered shortest path (LASH) algorithm [45]. The ALASH protocol improves the LASH layering function by considering the expected traffic patterns. We follow the priority layering function detailed in prior work [11]. Priority layering allocates as many virtual layers as possible to source-destination pairs with high traffic intensities. Additionally, for any source-destination pair, a path containing a wireless link is enabled only when using the wireless path gives rise to lower latency than taking the wireline-only path. More precisely, for each source-destination pair, the shortest path(s) are found first. If a path involving a wireless link is found to have a shorter length than the wireline-only path, this wireless path becomes an enabled path. These features improve message transfers under high traffic intensities by providing greater routing flexibility.

As we discuss later in Section 5.1, the CNN computations induce an asymmetric proportion of reply and request packets (i.e., the MC-to-GPU traffic volume is usually much higher than the GPU-to-MC traffic volume). In order to efficiently handle this asymmetric traffic, we require an efficient wireless MAC protocol that can dynamically allocate wireless bandwidth between MC-to-GPU and GPU-to-MC data transfers, depending on the current instantaneous communication requirements. For this purpose, we employ the distributed MAC protocol described in [44]. With this MAC, whenever a message wants to utilize the wireless channel, the state of the wireless medium is first checked. If the wireless medium is free, a wireless request period takes place. Each request period consists of *N* slots, a dedicated slot for each of the *N* WIs sharing the same wireless channel. During each request period, every WI sharing the same wireless channel has the opportunity to request the wireless medium by broadcasting to the wireless channel an "on" signal (**1**) in its allocated slot. An off-signal/no-transmission (**0**) in a particular slot would indicate that there is no request from the corresponding WI. For example, for a set of 4 WIs sharing a wireless channel, the request period has 4 slots (*b3 b2 b1 b0*). *b0* represents WI 0, *b1* represents WI 1, and so on. If WI 2 wants to request the wireless channel, WI 2 broadcasts a **1** during *b2* and listens for other requests during *b0, b1,* and *b3*. After the request period, all WIs follow a simple fairness-based common node selection algorithm [44]. Following the execution of this node selection algorithm, one of the requesting WIs acquires the wireless channel and starts transmitting data packets. When the wireless channel is busy, the packets are re-routed via the wireline links. Hence, it should be noted that, irrespective of their actual utilization values, the wireless links are inherently congestion-free and cannot become bandwidth bottlenecks; hence, the wireless channel utilization was not considered for the optimization formulation (Section 4.2.1).

| Parameters | Configuration |
|---|---|
| GPU Core clock | 1.5GHz |
| GPU Private L1 I cache | 64kB |
| GPU Private L1 D cache | 64kB |
| CPU Clock | 2.5GHz |
| CPU Private L1 I cache | 64kB |
| CPU Private L1 D cache | 64kB |
| Shared L2 cache size | 1MB per MC |
| DRAM | 3GB |

**Table 2. System configurations**

# 5 Experimental Results and Analysis

We utilize Gem5-gpu, a heterogeneous full system simulator to obtain processor- and network-level information [46]. Gem5-gpu is a combination of two well-known simulators: Gem5, a manycore CPU simulator and GPGPU-sim, a detailed GPGPU simulator. Gem5-gpu provides a customizable interconnection model through the Garnet network and supports inter-CPU-GPU cache coherence protocols [47].

We have modified this Garnet network topology to implement the WiHetNoC architecture. We employ Gem5-gpu in the full-system simulation mode with default task mapping/scheduling provided in Linux. The exact processing element composition can be optimized, but it is highly dependent on the specific deep learning architecture (*i.e.*, input data size, network depth and number of classes) and the software implementation of the deep learning framework. However, since the proposed optimization methodology uses traffic characteristics and knowledge of the CPU and GPU traffic requirements, our proposed methodology can be used for any composition of CPUs/GPUs/MC/s and system size. Hence, without loss of generality, in this work, we consider a heterogeneous architecture with 56 GPUs, 4 CPUs, and 4 MCs for our experimental analysis. This gives rise to a system size of 64 tiles, which are arranged in an 8 × 8 grid configuration. In the full-system simulations we consider NVIDIA Maxwell-based GPU cores and standard x86 CPU cores. We consider the MESI two-level cache coherence protocol. Each CPU and GPU streaming multiprocessor (SM) is allocated private L1 data and instruction caches. The considered memory system also incorporates four LLCs that are shared among all the CPUs and SMs. Table 2 provides detailed configurations of the architecture considered in this work. We limit the number of NoC links to be the same as a conventional mesh architecture so that our NoC does not introduce additional area overhead. We use GPUWattch to obtain detailed processor power profiles from the Gem5-gpu statistics [48].

We use the three-stage router microarchitecture for all NoCs under consideration. The delay of each stage is constrained within one NoC clock cycle (2.5 GHz). For routers with more than four inter-tile router ports, the output arbitration has an additional pipeline stage that takes one clock cycle. This is accounted for when determining the latency and energy of the proposed NoC architectures.

## 5.1 Analysis of On-Chip Traffic Patterns

In this section, we analyze the nature of the on-chip traffic patterns generated by training LeNet and CDBNet on the considered heterogeneous manycore platform

Since each layer of any CNN involves unique computation patterns and memory requirements, the volume of the on-chip traffic varies from one layer to another. Therefore, to capture the effects of these distinct traffic patterns, we will individually present the NoC-level experimental results for each convolutional (denoted as C), pooling (denoted as P) and fully-connected layer (denoted as F). We compare the network traffic intensity (measured in number of flits injected per second) associated with each neural network layer in Figs. 5(a) and (b). From these figures, it is clear that for both LeNet and CDBNet, and for both forward and backward passes, the convolutional layers exhibit high injection rates followed by pooling layer while the fully connected layer exhibits the lowest injection rate.

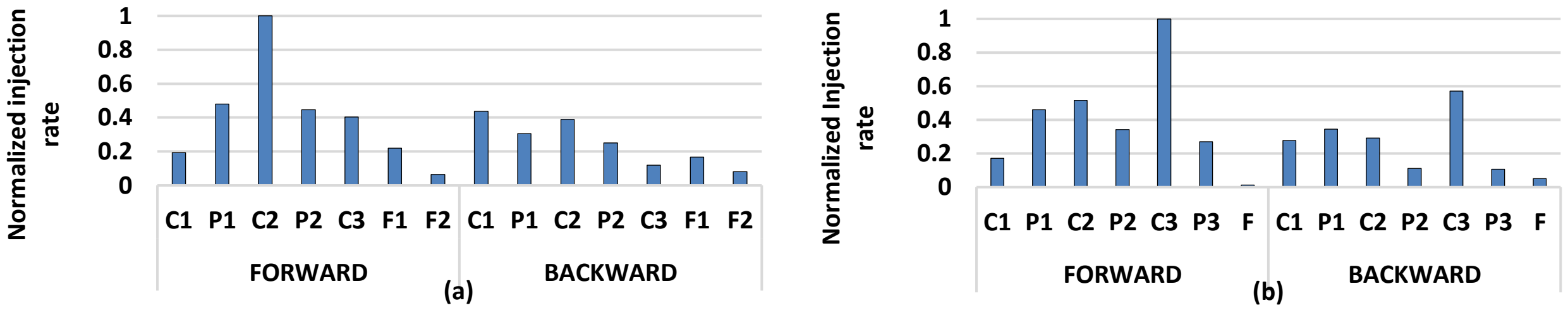

**Fig. 5. Message injection rate for training the (a) LeNet and (b) CDBNet (normalized with respect to the highest injection rate layer).**

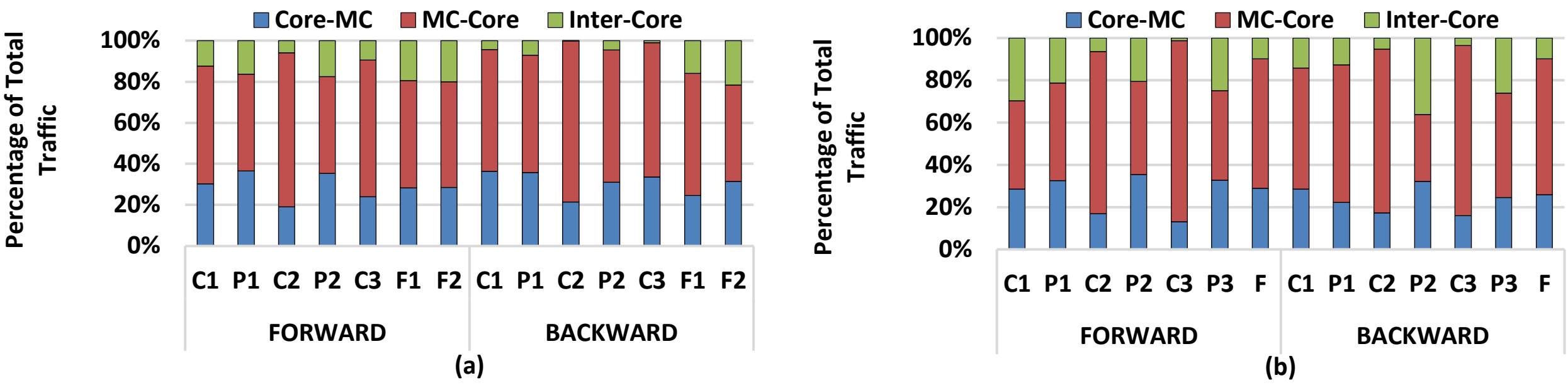

**Fig. 6 Traffic breakdown for each layer of the (a) LeNet and (b) CDBNet CNNs. The graphs aggregate the CPU and GPU traffic (collectively referred to as "Core") to emphasize the heavy many-to-few traffic and the asymmetric nature of the traffic to/from the MCs for training CNNs.**

Another important characteristic that we have observed is the asymmetric nature of the on-chip traffic patterns. The volume of GPU-to-MC traffic in heterogeneous systems is usually lower than the MC-to-GPU traffic volume [30]. Moreover, the CNN applications involve the execution of numerous consecutive compute-kernels[2]. Each compute-kernel involves a different ratio of input and output memory requirements as implied in Table 1. As it can be seen from Figs. 6(a) and (b), the variation in the input and output data volumes further adds to the asymmetry exhibited between the incoming and outgoing MC traffic for the CNN layers.

Apart from the similarity in injection rates and traffic asymmetry, we have also observed in Fig. 6 that the NoC traffic consistently follows the many-to-few, across all layers of the CNN computation for both LeNet and CDBNet architectures. The main source of the many-to-few traffic come from traffic moving to/from the MCs (the GPUs and CPUs greatly outnumber the MCs). This type of traffic makes up 93% for LeNet and 89% for CDBNet of the total traffic across all the CNN layers.

We also consider the temporal locality of the memory access for two representative CNN layers, namely, convolution and pooling, for LeNet (CDBNet exhibits similar behavior). It is clear from the Figs. 7(a) and (b) that in both layers many GPU cores transmit/receive data at the same time. In such situations, the network resources will be monopolized by GPU traffic, leading to significant degradation in CPU-MC communication latency. This demonstrates the need for single-hop shortcuts between the CPU and MC tiles. As mentioned in Section 4, we resolve this problem by employing dedicated wireless links between CPUs and MCs.

From these discussions, it is clear that both the considered CNN architectures give rise to very similar on-chip traffic patterns. Hence, by optimizing a NoC to efficiently handle the many-to-few and asymmetric nature of the on-chip data transfers, it is possible to design a single heterogeneous manycore architecture for implementing CNN computations.

### 5.2 Mesh NoC Link Utilization Characterization

In this section, we analyze the link utilization characteristics of the conventional wireline mesh NoC architecture

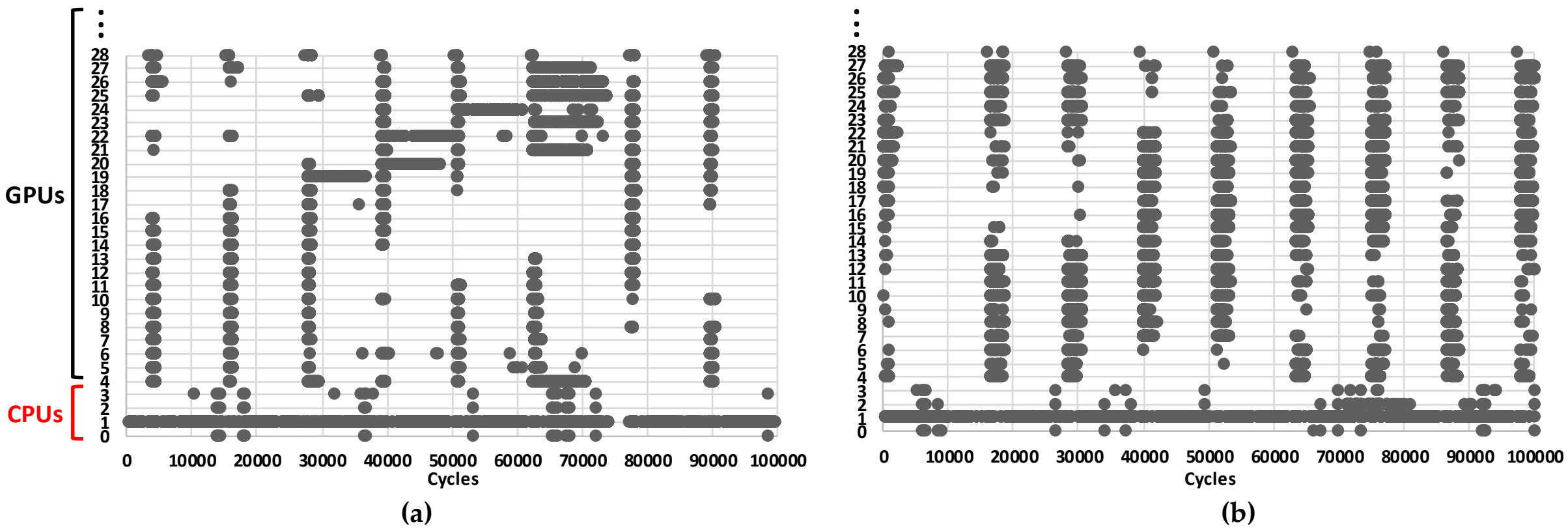

**Fig. 7. Snapshot of the memory access temporal locality during the forward pass of (a) convolution and (b) pooling layers running LeNet on the proposed heterogeneous architecture. Each dot represents receiving/transmitting data from/to the memory controllers. For instance, the dot at (10000, 3) indicates that CPU3 communicated with an MC at time 10000. Similar patterns are observed during subsequent iterations of these layers.**

[2] Each kernel corresponds to a particular CNN layer.

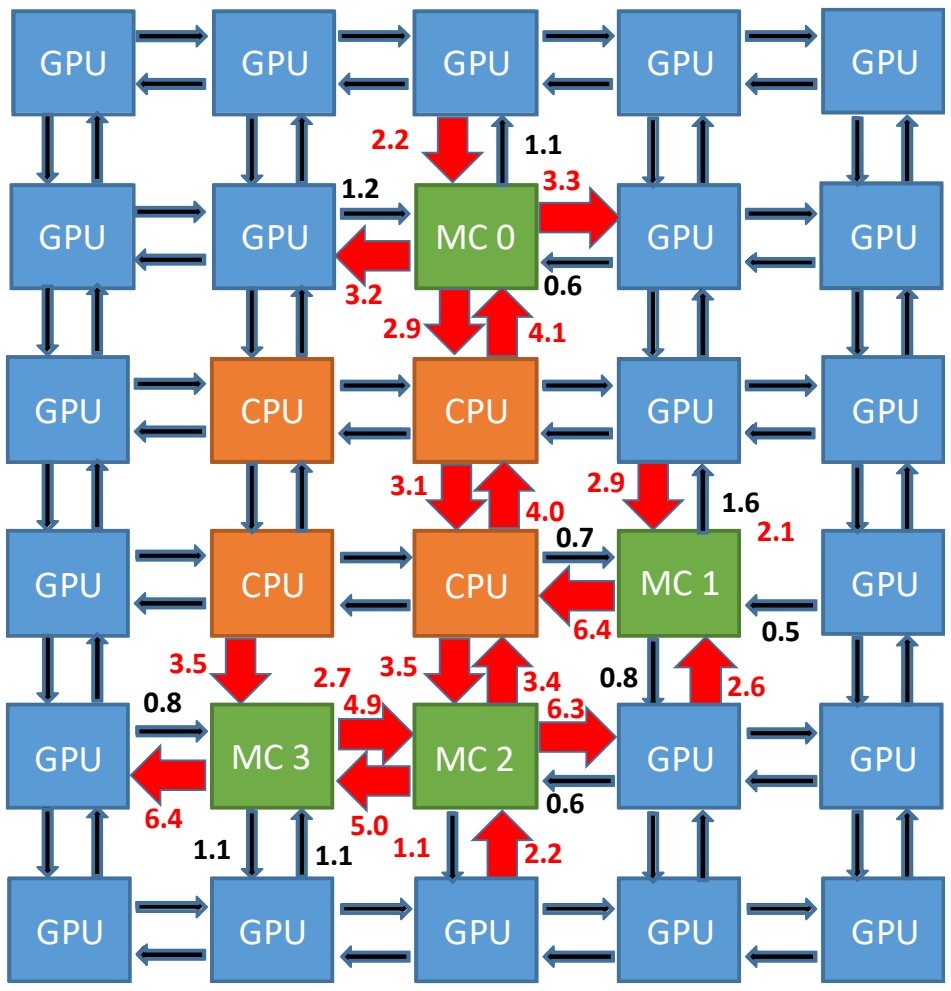

**Fig. 8. The central 6×5 portion of the optimized Mesh NoC indicating the link utilizations and the locations of CPUs and MCs for LeNet computations. All link utilizations are normalized with respect to the mean link utilization. The red arrows indicate the bandwidth bottlenecks whose utilization is at least 100% more than the mean.**

running the considered CNN applications.

Following the discussions in Section 4, for a mesh NoC with a given number of tiles, we first identify the positions of the CPUs and MCs such that the CPU-MC communication latency is minimized while the overall NoC throughput is maximized to enable efficient GPU-MC data exchange (under the *many-to-few* communication pattern). Following [49], we employ the AMOSA algorithm (mentioned in Section 4.2.2) to determine the optimal positions of CPUs, GPUs, and MCs in a mesh NoC to jointly optimize CPU-MC communication latency and the overall NoC throughput. Fig. 8 shows the core types and interesting link utilization behaviors in the central section of the final optimized mesh NoC. All link utilizations are normalized with respect to the mean link utilization. We can observe from Fig. 8 that in the optimized mesh NoC, the MCs and CPUs are clustered in the middle so that the CPU-MC communication latency is minimized. Moreover, placing the MCs closer to the middle ensures that handling of GPU-MC traffic is distributed among multiple MC ports, leading to higher NoC throughput.

Also, in Fig. 8, it is evident that even in this optimized NoC, a few links are more heavily utilized when compared to the rest of the links. In a previous work [49], it was demonstrated that links associated with MCs have nearly 500% higher traffic density than the overall average link utilization for the Rodinia backpropagation benchmark [50]. However, this backpropagation benchmark is much simpler than the workloads addressed in this work and only employs a single NN layer. For the LeNet and CDBNet computations considered in this work, the incoming vertical and outgoing horizontal links associated with the MCs have up to 600% and 700% higher traffic densities than the overall average respectively. These results can be attributed to the fact that the on-chip communications involved with the LeNet and CDBNet applications exhibit a high degree of asymmetric traffic and is many-to-few in nature (Figs. 6 and 8). When such traffic is handled using the traditional XY routing mechanism in mesh NoCs, the link utilizations become highly skewed.

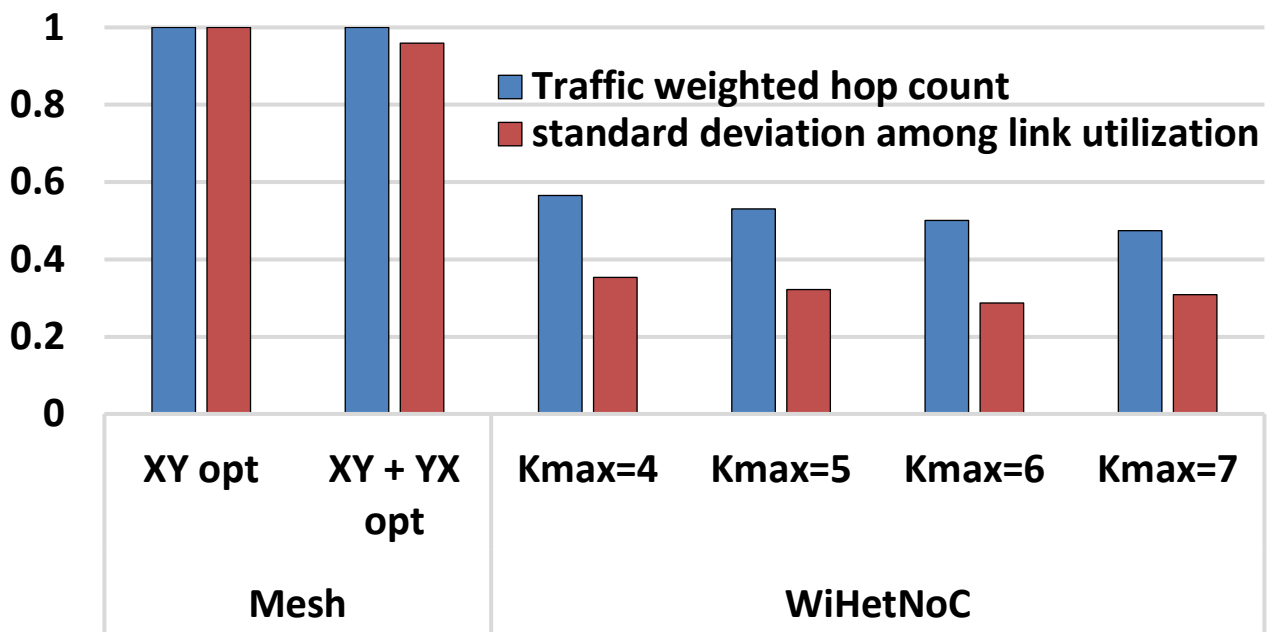

**Fig. 9. Traffic-weighted hop count and standard deviation among the link utilizations for the optimized mesh (denoted as OPT) and WiHetNoCs (for each considered $K_{max}$). For mesh NoCs we show the results for both XY and XY+YX routing schemes.**

In order to alleviate the traffic congestion caused by aggregation of traffic along the mesh NoC links, one can also adopt a combination of minimal XY and minimal YX routing as proposed in [29]. However, such an approach cannot eliminate the bandwidth bottlenecks from the optimized mesh NoC under the heterogeneous computing induced traffic patterns. To elaborate more, Fig. 9 shows the average traffic-weighted hop-count and the standard deviation among the link utilizations for the following NoC configurations: optimized mesh and four different WiHetNoC candidate architectures (these WiHetNoC candidates are explained later in detail). As it can be observed from Fig. 9, when compared with WiHetNoC candidates, both the standard deviation and the traffic-weighted hop-count of the optimized mesh NoC are at least 2× higher. This indicates the presence of traffic hotspots that can lead to bandwidth bottlenecks.

It can be observed from Fig. 8 that in the optimized mesh NoC, since the MCs are placed around the CPU cluster, the CPU tile routers also act as intermediate routers forwarding the GPU-MC traffic. With continuous streams of data flowing from MCs towards GPUs, the above-mentioned traffic forwarding causes traffic congestion in the CPU tile routers, leading to an undesired increase in the CPU-MC communication latency. To resolve this issue, the MCs can be placed away from the CPUs such that the CPU tiles are not involved in the GPU-MC traffic forwarding. However, due to the lack of long-range shortcuts in mesh NoC, placing the MCs away from the CPUs increases the number of hops required for CPU-MC communication leading to network latency penalties. Hence, our target is to design an application-specific heterogeneous NoC architecture with suitable long-range shortcuts to improve both CPU-MC latency and the overall NoC throughput. Since the wireless links make the CPU-MC latency agnostic of their placements, we keep the CPUs at the center of the system and distribute the four MCs to the center tiles in each of the four quadrants of the system. All the other tiles are occupied by the GPU cores surrounding the CPUs and MCs. The whole system is then integrated using the WiHetNoC framework. In the following section, we determine the various network parameters of this WiHetNoC.

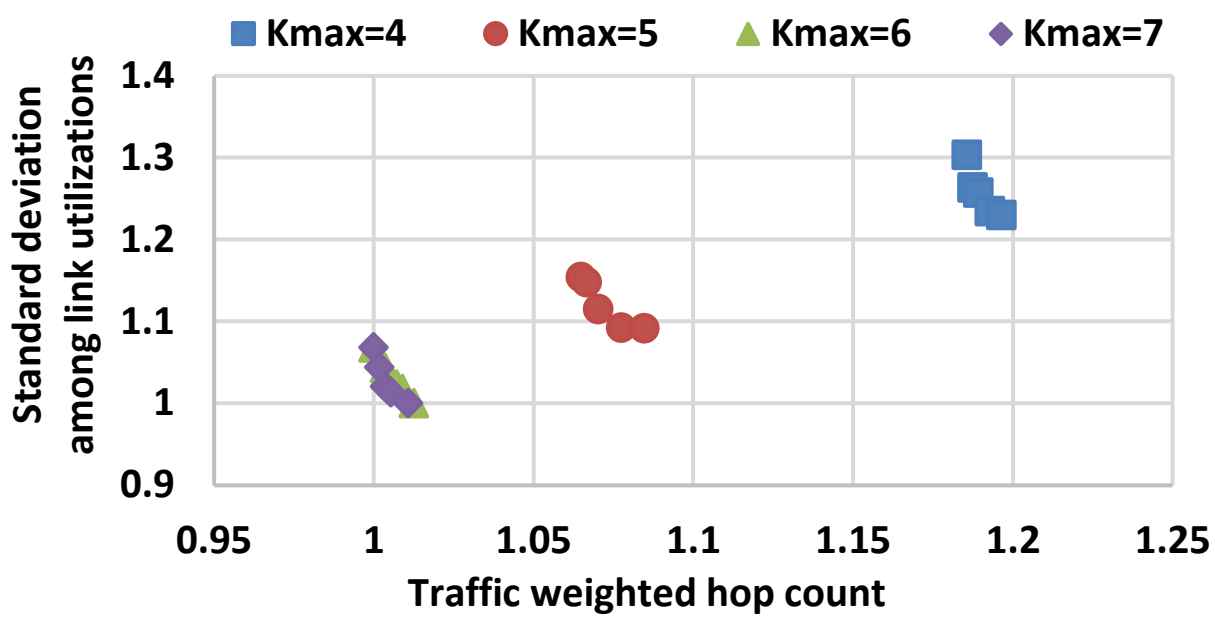

**Fig. 10. Normalized traffic-weighted hop count and standard deviation among the link utilizations of various candidate wireline configurations for WiHetNoC ($K_{max}$ values ranging from 4 to 7).**

## 5.3 Determining WiHetNoC Parameters

In this section, we determine the overall architecture and the network parameters of WiHetNoC.

### *5.3.1 Router Port Upper Bound*

In the WiHetNoC design process, we first find the optimum value of $k_{max}$ (maximum number of inter-tile communication ports in a router). As we discussed in Section 4.2.2, we start with a $k_{max}$ range of 4 to 7. For each $k_{max}$ value we use the above-mentioned AMOSA optimization to find a final candidate solution set. Fig. 10 shows the mean link utilization ($\bar{U}$) and the standard deviation among the link utilizations ($\sigma$) for all the solutions in the final candidate set corresponding to each $k_{max}$ value. All the values in this figure are normalized with respect to the final WiHetNoC configuration. As it can be observed from the graph, as the $k_{max}$ value is increased, the $\bar{U}$ and $\sigma$ values decrease. This happens because for higher $k_{max}$ values, more inter-router connections are allowed in each MC, leading to lower average hop counts between GPUs and the MCs. However, Fig. 10 shows diminishing gains in hop count and link utilization standard deviation reductions with increasing $k_{max}$ and there is no gain in exploring beyond $k_{max} = 7$. As explained in Section 4.2.2, the final WiHetNoC wireline connectivity is identified from the candidate networks by comparing their network Energy-Delay-Products (EDP). Average message latency and energy are used in this EDP computation.  Fig. 11 shows the EDPs of the optimal networks corresponding to each $k_{max}$, (thus a total of four optimal networks are shown corresponding to $k_{max}$ values ranging from 4 to 7). From Fig. 11 it is evident that the optimal value for $k_{max}$ is 6. Beyond this value of $k_{max}$, the EDP worsens due to higher router energy consumptions without significant gains in network latency and hop count. In addition to our comments in Section 5.2, Fig. 9 demonstrates the network characteristics for the four optimal WiHetNoC architectures (for the four considered $k_{max}$ values), showing an inflection point specifically for the standard deviation at $k_{max} = 6$. Hence, we select the network corresponding to $k_{max} = 6$ as the optimal wireline connectivity for WiHetNoC.

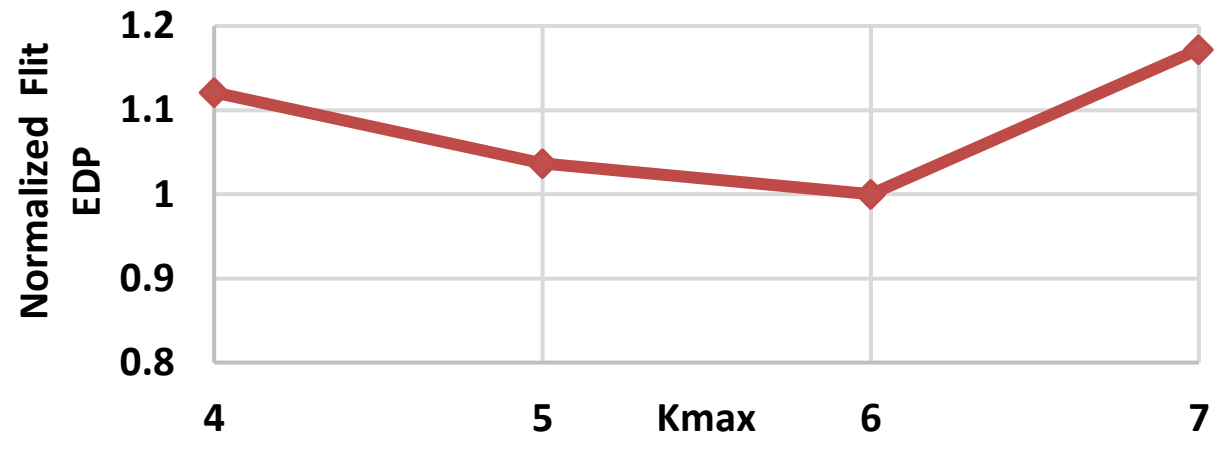

**Fig. 11. Variation in network EDP for different router port upper bounds ($K_{max}$).**

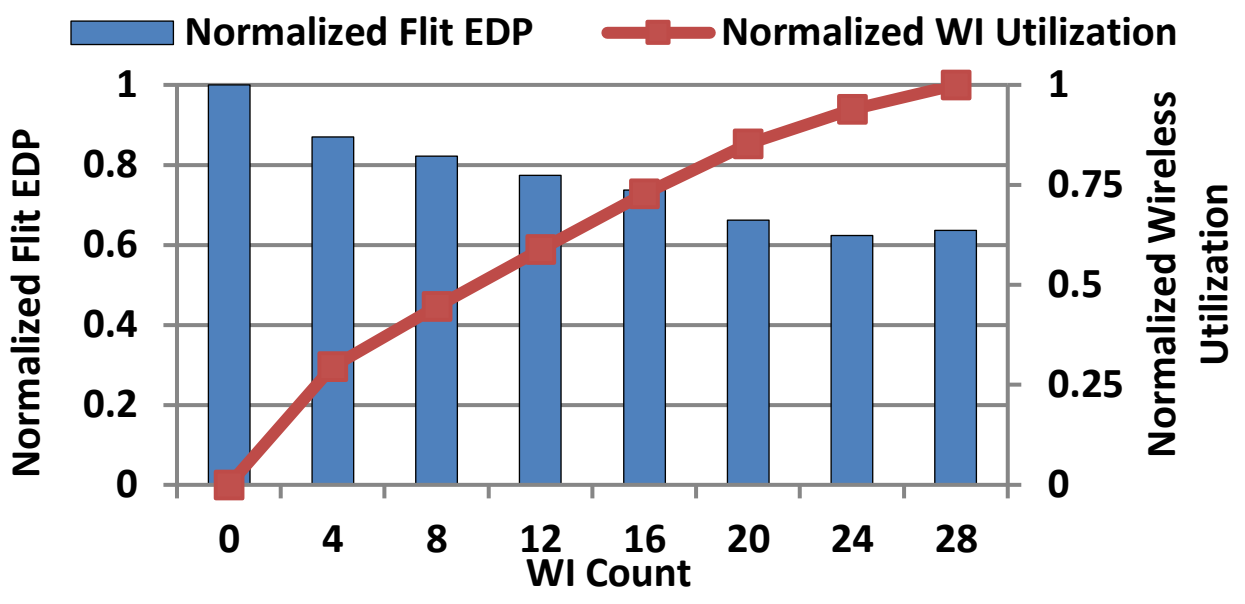

**Fig. 12. EDP and Wireless utilization for various WI counts.**

### *5.3.2 Number of WIs*

Next, we identify the number of WIs needed for the GPU-MC communication in the WiHetNoC architecture. As mentioned earlier, we are able to create five non-overlapping channels. We dedicate one channel to achieve single-hop CPU-MC communications. Hence, four wireless channels are available for GPU-MC communication.

Fig. 12 shows the variation in EDP and wireless utilization observed with varying WI counts. The wireless utilization parameter represents the percentage of total messages that are using the wireless channels. As observed from this figure, the EDP initially reduces as the WI count increases: higher number of wireless shortcuts improves the wireless utilization and hence, lowers the overall network latency. However, beyond a WI count of 24, with more than six WIs allocated on a single wireless channel, the MAC overhead (and hence, the channel access latency) and, therefore, the network EDP starts to increase [44]. Hence, in our WiHetNoC, we employ 24 WIs for GPU-MC communication (four channels are used with six WIs operating on each channel). Fig. 13 shows the effects of adding wireless channels for GPU-MC communication on the performance of the WiHetNoC. With increasing number of wireless channels, the amount of data using wireless medium increases and the overall EDP improves. However, the enhancement in wireless utilization and subsequent improvement in EDP slows down beyond a certain number of wireless channels. For the 64-tile system size, increasing the number of wireless channels beyond 4 does not enhance the system performance noticeably as the opportunity for more wireless utilization diminishes.

Since each WI transceiver occupies an area of 0.25mm$^2$,

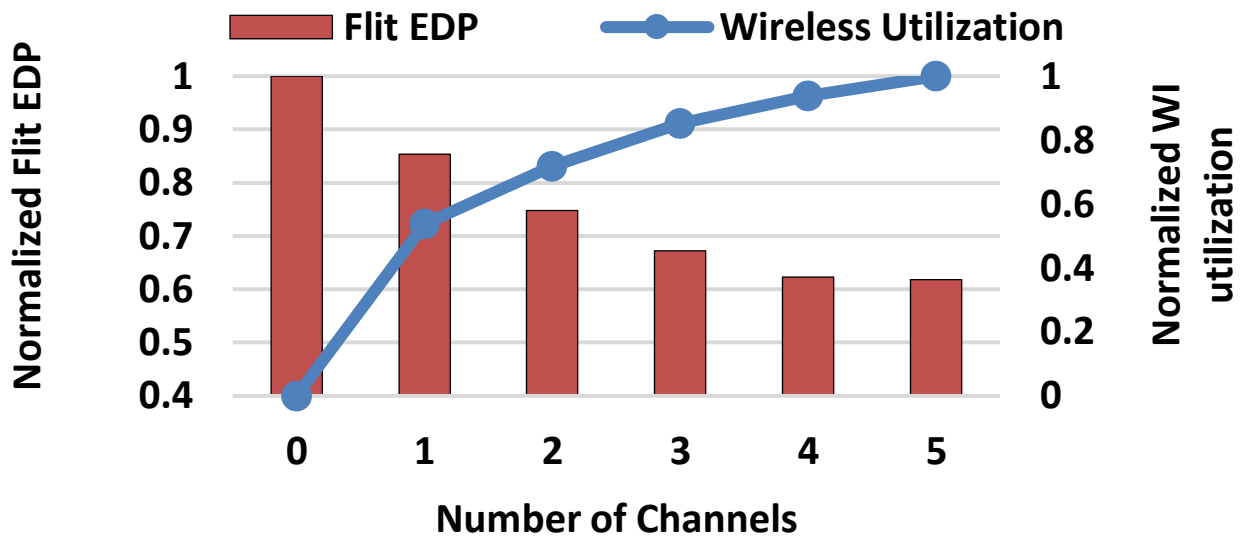

**Fig. 13. EDP and WI utilization with various number of channels.**

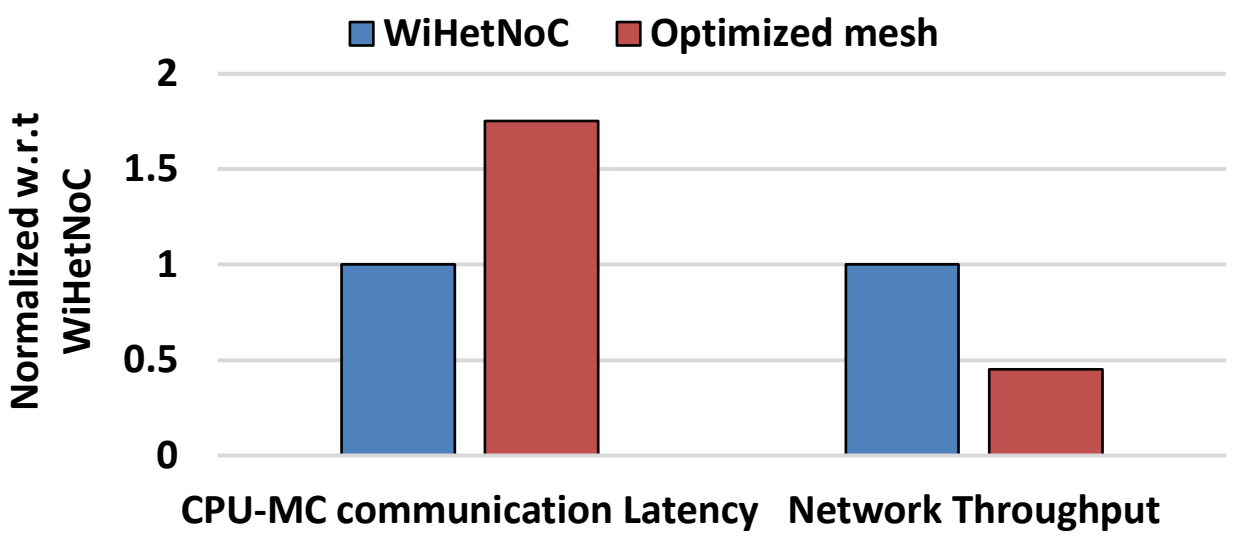

**Fig. 14. CPU-MC communication latency and overall NoC throughput for optimized mesh NoC and the optimized WiHetNoC.**

the addition of 24 WIs introduce a total of 1.82% silicon area overhead for a die with dimensions of 20 mm×20mm.

### *5.3.3 Characteristics of WiHetNoC*

Fig. 14 compares the CPU-MC communication latency and overall network throughput achieved with the optimized WiHetNoC against the optimized mesh (explained earlier in Section 5.1). From this graph, we can observe that WiHetNoC achieves both higher NoC throughput and lower CPU-MC communication latency than the optimized mesh NoC (while requiring a similar implementation overhead). The WiHetNoC improves the throughput by a factor of two compared to the optimized mesh configuration. This demonstrates the effectiveness of the proposed methodology in designing NoC architectures for heterogeneous platforms with many-to-few communication patterns.

In Fig. 15, we show the Cumulative Distribution Function (CDF) of the link utilizations for the WiHetNoC and the optimized mesh NoC with XY+YX routing scheme (denoted as Mesh_opt). The utilizations in this figure are normalized with respect to the mean link utilization observed in Mesh_opt ($U = 1$ represents this mean utilization). It is clear from Fig. 15 that, 20% of the Mesh_opt NoC links have at least 2x higher utilization compared to the mean utilization. In addition, more than 90% of WiHetNoC links fall under the mean link utilization of the mesh NoC. Generally, when compared to Mesh_opt NoC, the WiHetNoC CDF curve is shifted left indicating a reduction in the overall link utilizations, which is obtained through lowered inter-router hop counts. Moreover, as shown in Fig. 15, unlike Mesh_opt, WiHetNoC has no links with very high utilizations (no links with $U > 2$). Thus, WiHetNoC greatly reduces bandwidth bottlenecks.

Aside from avoiding the wireline bandwidth bottlenecks, the WiHetNoC is also highly suitable in handling the asymmetric traffic patterns that arise from heterogeneous CNN training computation. By using the distributed

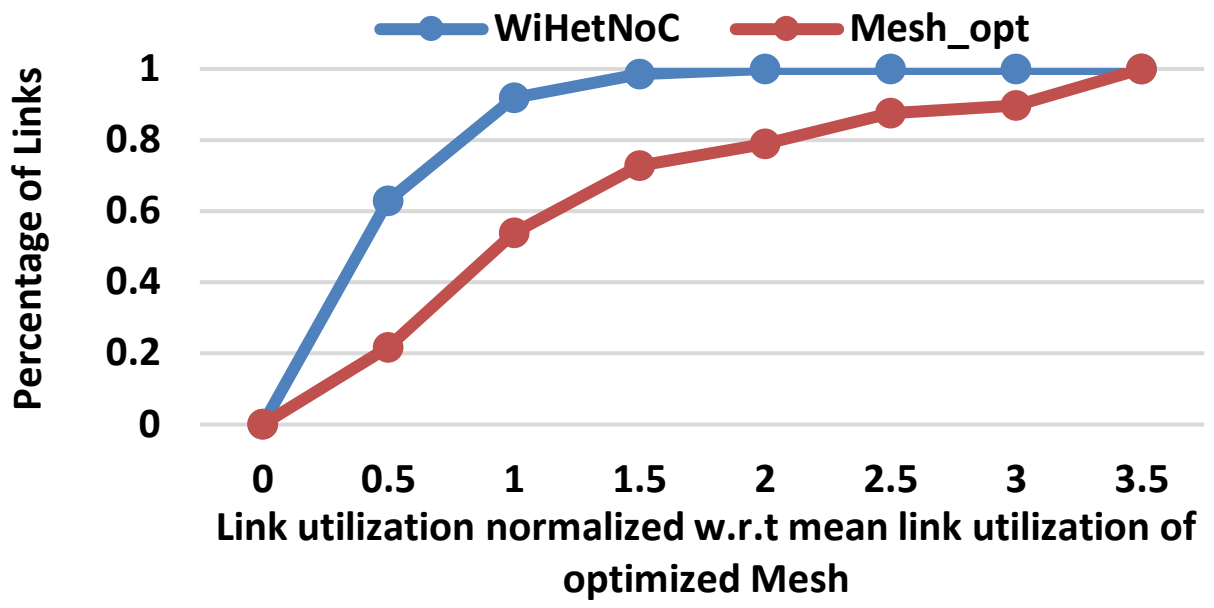

**Fig. 15. Cumulative Distribution Function (CDF) of link utilizations for the WiHetNoC, and optimized mesh architectures.**

MAC protocol (explained in Section 4.2.5), the WiHetNoC enables efficient dynamic allocation of wireless bandwidth for MC-to-Core (both CPU and GPU cores) and Core-to-MC data transfers, depending on the instantaneous communication requirements. This is further corroborated by Figs. 16 (a) and (b), which compare the ratio between the MC-to-Core and Core-to-MC data transfers that utilize the on-chip wireless interfaces. It is evident that the wireless utilization aligns well with the asymmetry ratios shown earlier in Fig. 6. This proves that the WiHetNoC is inherently capable of performing efficient dynamic bandwidth allocations as per the communication requirements.

## 5.4 NoC Performance Evaluation

In this section, we compare the performance of the WiHetNoC and the optimized mesh NoC for running the CNN training applications considered in this work. In this comparative performance evaluation, we also consider the HetNoC: a wireline-only architecture that replaces all of WiHetNoC's wireless links with pipelined long-range metal wires.

We first present the network-level analysis showing both the latency and message EDP (energy delay product per message). Figs. 17 (a-b) and 18 (a-b) show the network latency and EDP, respectively, of WiHetNoC and HetNoC, with respect to the optimized mesh configuration for each layer of LeNet and CDBNet. For LeNet, on an average, the HetNoC reduces the network latency and EDP by **23**% (**22**% for CDBNet) and **44**% (**42**% for CDBNet) respectively, when compared to the mesh NoC. With the use of long-range shortcuts between physically remote nodes, the HetNoC enables a lower average hop count than the optimized mesh NoC, and hence, achieves significant reductions in intermediate flit counts. However, as we stated in Section 4.2.3, the long wireline links of the HetNoC suffer from high link latency and energy consumption. In the WiHetNoC architecture, many of these long-range wireline links are replaced with energy-efficient wireless links, and hence, WiHetNoC enables **24**% more latency improvement for both LeNet and CDBNet and **30**% and **28**% more EDP improvement in average for LeNet and CDBNet respectively, when compared to the HetNoC. From Fig. 17 (a-b), it is evident that the WiHetNoC achieves **41.8**% and **42**% lower network latency on an average for LeNet and CDBNet respectively compared to the mesh architecture. In addition, as shown in Fig.18 (a-b), WiHetNoC also saves the EDP by **60**% and **58**% on an average for LeNet and CDBNet respectively, compared to the same mesh NoC.

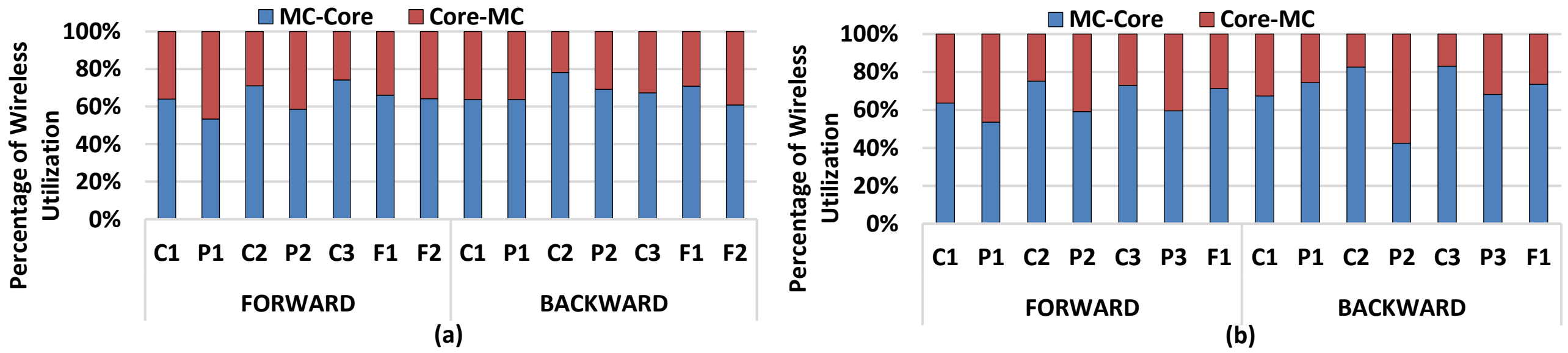

**Fig. 16. Asymmetric nature of WI utilization for training the (a) LeNet and (b) CDBNet.**

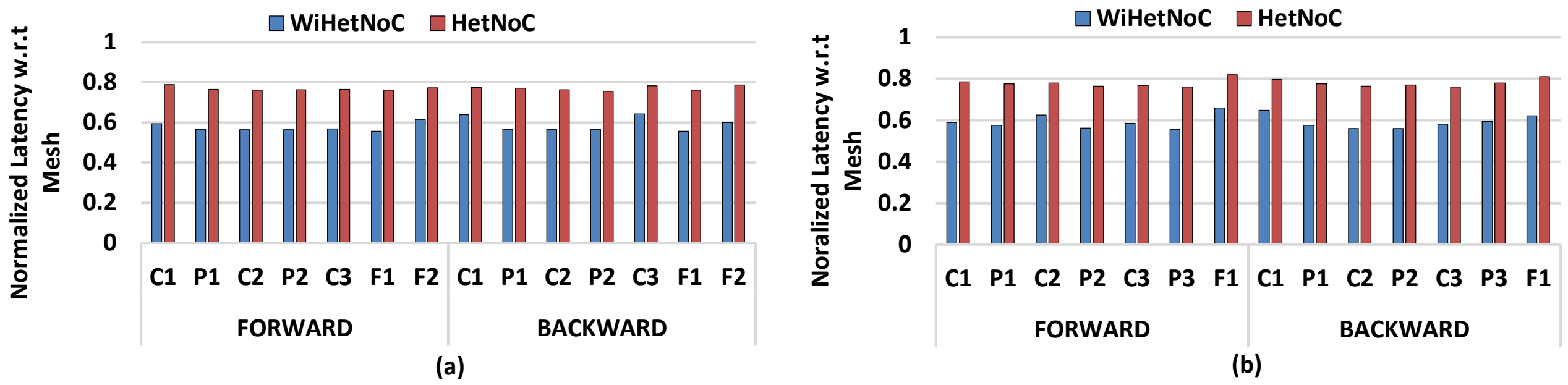

**Fig. 17. Normalized network latency for training the (a) LeNet and (b) CDBNet.**

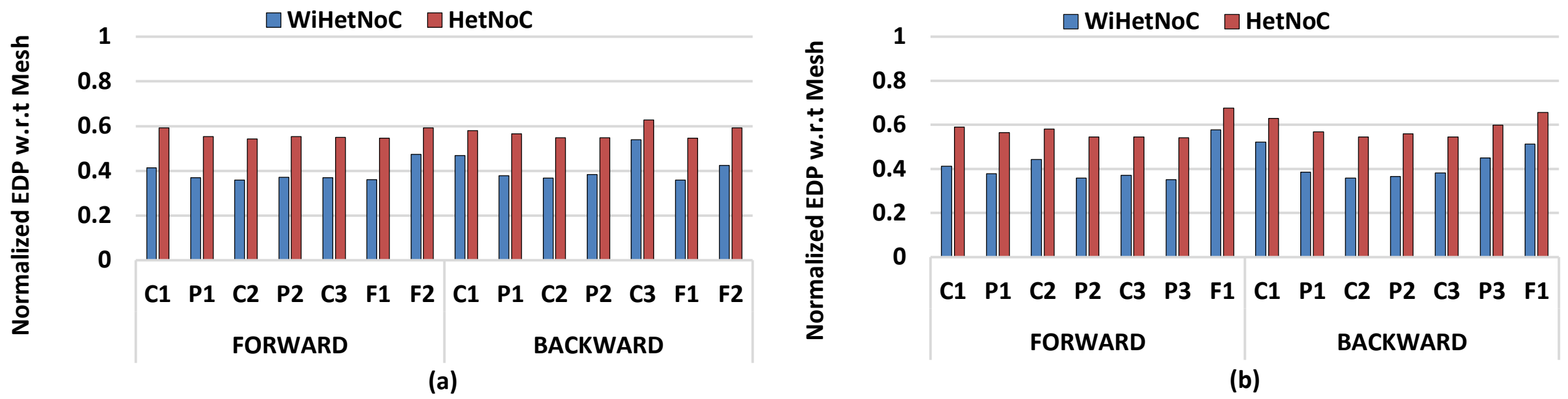

**Fig. 18. Normalized network EDP for training the (a) LeNet and (b) CDBNet.**

From Figs. 17(a-b) and 18(a-b) we can also observe that the achieved WiHetNoC gain varies from one CNN layer to another. Layers that exhibit high traffic intensities (layers P1, C2 and P2 in Fig. 5(a)) benefit more from the advantages provided by the WiHetNoC and hence achieve more latency and EDP improvement (over mesh in Figs. 17 and 18) than the layers that involve low injection rates (layers F1 and F2 in Fig. 5(a)). The fully connected layers show the lowest injection rate, and most of the NoC traffic in this layer is exchanged between CPU and MCs. For this case, the optimized mesh can handle the CPU-MC traffic without creating noticeable bottleneck as MCs are placed closer to the CPUs when compared to CNN layers with higher injection rates. Therefore, as shown in Figs. 17 (a-b) and 18 (a-b), the achievable performance gain of WiHetNoC is lower for the fully connected layers when compared to that of the convolutional layers.

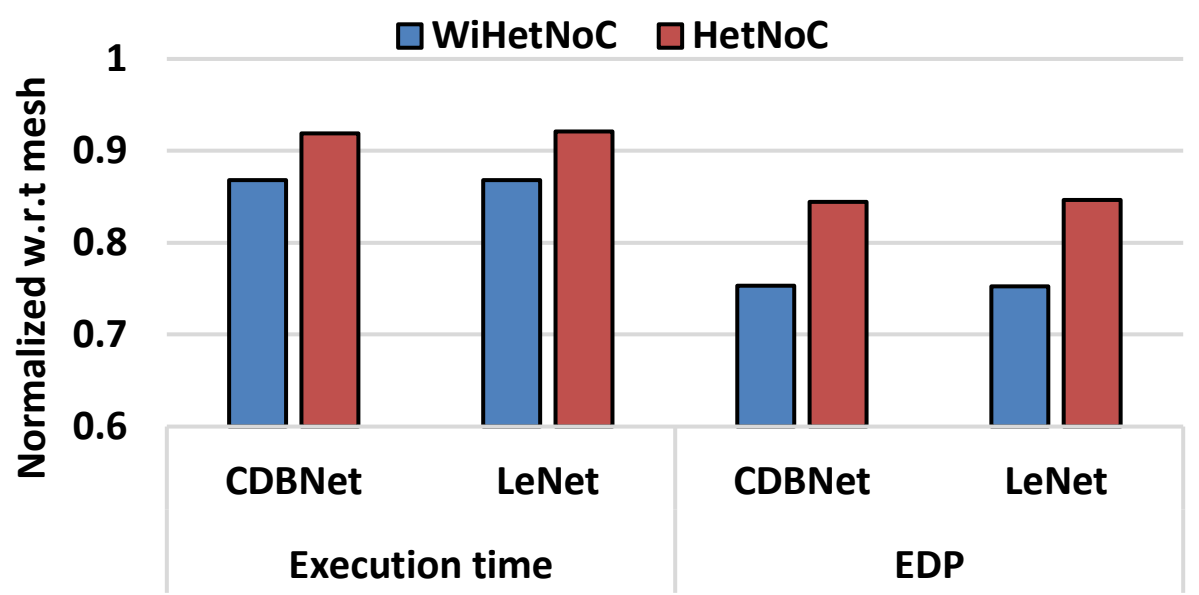

**Fig. 19. Normalized execution time and full system EDP for training LeNet and CDBNet.**

### 5.5 Full-System Performance Evaluation

In this subsection, we compare the overall application execution time and the full-system EDP for the optimized mesh, HetNoC, and WiHetNoC interconnected manycore platforms. Fig. 19 shows the execution time and full-system EDP for WiHetNoC and HetNoC, with respect to the optimized mesh architecture for both CNN training applications. The HetNoC achieves **8.1**% and **8**% execution time improvement for CDBNet and LeNet respectively over the mesh while the WiHetNoC shows **13.1**% and **13.2**% improvement for CDBNet and LeNet respectively over the mesh NoC. The dedicated wireless channel for CPUs on the WiHetNoC enables a highly efficient data transfer between CPUs and MCs. Also, the use of wireless shortcuts helps in achieving high-bandwidth and low-latency GPU-MC communications. These benefits translate to a **25**% full-system EDP reduction with WiHetNoC for both CDBNet and LeNet, when compared to the optimized mesh NoC.

## 6 Conclusions and Future Work

The computation patterns arising from training CNNs can be efficiently handled by a single-chip heterogeneous manycore platform. Conventional NoC architectures, such as mesh, are unable to handle the contrasting communication requirements imposed by CPUs and GPUs.

In this paper, we have proposed the design of a hybrid

NoC-enabled single-chip heterogeneous computing platform for energy-efficient acceleration of CNN training. By virtue of using single-hop wireless links, the proposed WiHetNoC architecture achieves a much better GPU-L2 throughput and a lower CPU-L2 communication latency, when compared to a highly-optimized mesh NoC. Thus, the proposed NoC architecture is able to efficiently fulfill the communication requirements of *both* CPU and GPU cores. For the considered CNN applications, WiHetNoC achieves **25**% lower full system EDP with respect to the mesh and **15**% lower full system EDP when compared to a fully wireline application-specific architecture.

It should be noted that this full system EDP improvement comes only from the network level innovation. The full system EDP will improve further when the WiHetNoC design is complemented with suitable core-level task and power management strategies, which will be the focus of our future investigation.